\documentclass[11pt,a4paper]{article}

\usepackage{amssymb,amsmath} 
\usepackage{graphicx}
\usepackage{subfig}
\usepackage{epsfig}

\usepackage{cite}
\usepackage{multicol}
\usepackage{color}
\usepackage{framed}
\usepackage{textcomp}
\usepackage{stmaryrd} 

\usepackage[hmargin=2.5cm, vmargin=3.5cm]{geometry}
\usepackage{enumerate}

\usepackage{bm} 
\usepackage{amsmath}
\usepackage{amsthm} 
\usepackage{amstext}
\usepackage{mathrsfs} 
\usepackage{pifont} 

\usepackage{url} 
\newcommand{\kommentar}[1]{}

\newcommand{\Real}{\mathbb{R}}

\newcommand{\Nat}{\mathbb{N}}

\theoremstyle{definition}

\newcommand{\gausslower}[1]{\left\lfloor #1 \right\rfloor}

\renewcommand{\d}{\,\text{d}}

\renewcommand{\d}{\,\text{d}}

\newcommand{\ONotSym}{\mathcal{O}}

\newcommand{\keywords}[1]{\textbf{Keywords:} #1}

\newcommand{\myheading}[1]{\vspace{-0.2cm}\mbox{ }\\ \textbf{#1.}}

\begin{document}

\title{Deterministic bootstrapping for a class of bootstrap methods}

\author{Thomas Pitschel\footnote{Correspondence address: th.pitschel at tu-braunschweig dot de}}
\date{March 26, 2019}
\maketitle

\begin{abstract}
An algorithm is described that enables efficient deterministic approximate computation of the 
bootstrap distribution for any \emph{linear} bootstrap method $T_n^*$, alleviating
the need for repeated resampling from observations (resp. input-derived data). 
In essence, the algorithm computes the distribution function from a linear mixture of
independent random variables each having a finite discrete distribution.
The algorithm is applicable to elementary bootstrap scenarios (targetting the
mean as parameter of interest), for block bootstrap, as well as for certain 
residual bootstrap scenarios. Moreover, the algorithm promises a much
broader applicability, in non-bootstrapped hypothesis testing.

\end{abstract}

\mbox{ }\\

\myheading{Problem setting, motivation, related work}
Given an estimation problem based on real-valued data points $X_i$, $i=1\dots n$, deemed
to originate from some distribution $P^{\otimes n}$, and aiming to infer 
a parameter $\theta \in \Real^m$ of that distribution using a map $\hat\theta: (X_1, \dots, X_n) \mapsto \Real^m$,
the question arising in practice is with what certainty the obtained value
is close to the true value, or similarly what is the suitable confidence 
band (/volume) in which the true value can be expected to be, given the 
obtained estimate $\hat \theta = \hat\theta(X_1, \dots, X_n)$ and some confidence 
level.

As far as the joint distribution $P^{\otimes n}$ of the $X_i$ is known in parametric form, such intervals can be determined 
for every confidence level $0 < \alpha < 1$ by examining the then (in principle) known limit distribution
of a proxy quantity $T(X_1, \dots, X_n)$ which essentially is a suitably centered and normed ''derivate'' of 
$\hat \theta(\cdot)$. In absence of such
information, but under certain conditions, bootstrap methods allow to still 
reasonably estimate confidence intervals by determining an empirical distribution
of the proxy quantity for the parameter estimator of interest. To this end, bootstrap methods
resample from the existing data set, thereby generating ''replicate'' data sets, and 
determine an estimate for each replicate.

A common way to construct a bootstrap estimator is to use the original
$T$, replace the expectation parameter contained for centering by the empirical mean of
the data set
and apply it on data points $X_i^*$ which are obtained by resampling the original
data set. 
The so constructed bootstrap estimator $T_n^*$ is called plugin estimator (associated with $T$). The estimator and the 
exact specification of choosing the $X_i^*$ from the $X_i$ together constitute a bootstrap method.

This present text is concerned with \emph{linear bootstrap methods}, i.e. a class of methods where 
the bootstrap estimate $T_n^*$ belonging to the value of interest can be expressed linearly 
in the bootstrap variables, 
and the bootstrap variables are (conditional on the data set) 
stochastically independent through choice of the resampling rule. A list
of examples of such methods are found in the appendix \ref{sec:ExamplesOfLinearBootstrapMethods}.

The standard approach of constructing an empirical distribution of
$T_n^*$ would evaluate the map $T_n^*$ at various, say $B \in \Nat$, bootstrap samples, each of which is obtained
by resampling. The $B$ may not be chosen too small in order to ensure a sufficient
convergence of the distribution estimate to the actual distribution of $T_n^*$.
Since each bootstrap estimate evaluation naively needs at least $n$ data access operations, 
this approach has a computationally substantial cost for large $n$, concretely
is of complexity at least $\Omega(B \cdot n)$.

The claim here made is that for linear bootstrap methods, a much more direct 
method for obtaining an approximation to the distribution of $T_n^*$ may be employed.
Essentially, it is sufficient to approximate the discrete distribution of a linear mixture
of independent random variables which themselves are finitely discretely distributed.
The present text outlines a sketch of a suitable algorithm for this setting.
In effect, by doing away with the resampling in case of linear bootstrap estimators, 
the accuracy of the obtained distribution is substantially increased, since clearly any 
effect from variance introduced at the resampling level is avoided. (There have been previous attempts
to counter this variance, for example, by using a balanced bootstrap, see \cite{DavisonHinkleySchechtmann1986EfficientBootstrapSimulation}.
The principle behind the balanced bootstrap is to control the choice of the samples across many replicates so as to produce
a bootstrap mean estimate which is zero. Besides the drawback that this and similar methods
retain considerable simulation induced variance, one has to observe that the estimates
coming out of the generated single replicates are not strictly independent; ''later drawn'' replicates
chose values of their bootstrap variables not according to an Efron distribution.
Another, logically natural, approach for ''stabilizing'' the outcome of bootstrap simulations 
is to use quasi-Monte Carlo methods in an attempt to cover the simulation space more evenly
and thus remove some of the unwanted randomness of the simulation procedure. This approach
has been followed for example by Kolenikov \cite{RePEc:boc:asug07:11}, Aidara \cite{Aidara2013BootstrapVarianceEstimationForComplexSurveyData}
and references therein.)

It is noteworthy that the method here described extends 
to certain non-linear
mixtures as well, namely $Z = \sum_{j=1}^m a_j h_j(X^{[j]})$ with $X^{[j]}$ independent, and furthermore
is cascadable. Therefore random variables of form $Z = Y^{[0]} + (\sum_{j=1}^m a_j Y^{[j]})^2$, with $Y^{[j]}, j=0\dots m,$ 
independent are amenable to the proposed method.

The remainder of the text first gives a description of the algorithm, and then comments on 
the output obtained from a small (synthetic) numerical example.

\myheading{Algorithm description}
Let $a_j \in \Real$, $j=1\dots m$, and let a finite discrete distribution $\hat F_n$ of a random variable
be given in form of real values $X_i$, $i=1\dots n$. Let $Z = \sum_{j=1}^m a_j X^{[j]}$,
where $X^{[j]}$ are independently distributed according to the distribution $\hat F_n$.
Choose $N \geq n$, and choose $T \geq T_Z$ with $T_Z := z_U - z_L$ and
\begin{align}
    z_U & = \max_i X_i \cdot \sum_{a_j > 0} a_j  +  \min_i X_i \cdot \sum_{a_j < 0} a_j \\
    z_L & = \max_i X_i \cdot \sum_{a_j < 0} a_j  +  \min_i X_i \cdot \sum_{a_j > 0} a_j.
\end{align}
An approximation of the probability density $f_Z$ is computed as follows:

\mbox{ }\\
1. Compute\footnote{In this and later expressions, it denotes $i$ after $2 \pi$ the imaginary unit.} for all $k=0 \dots (N-1)$ and all $j=1\dots m$,
\begin{align}
    g_{k,j} = \frac{1}{n} \sum_{i=1}^n e^{-2 \pi i \cdot \frac{a_j X_i}{T_Z} \cdot k}.
\end{align}
2. Set
\begin{align}
    g_{k} = \prod_{j=1}^m g_{k,j}.
\end{align}
3. Using an inverse Fast-Fourier transform, compute for $i=0 \dots (N-1)$
\begin{align}
    \tilde f_i = \frac{1}{N} \sum_{k=0}^{N-1} g_k e^{2 \pi i \cdot \tfrac{i k}{N}}. \label{eqn:algo_InvFourierTrafo}
\end{align}
4. Set $f_i := 2 \text{Re}(\tilde f_i) - \frac{1}{N}$.

\vspace{-2mm}
\mbox{ }\\
Then, under suitable conditions, $f_i$ approximates 
$\int_{z_0}^{z_0 + T_Z/N} f_Z(z) \d z$ with $z_0 = i \frac{T_Z}{N} - T_Z \cdot 1_{\{i \frac{T_Z}{N} > z_U\}}$.
Consequently, it is $h_i = \sum_{i'=0}^i f_{i + i_{min}}$ with $i_{min} = \gausslower{z_L (N/T)} \text{ mod } N$ approximating
$F_Z(i \cdot \tfrac{T_Z}{N} + z_L)$ for $i=0 \dots (N-1)$.
Note that, given a sample, the computation of quantiles is deterministic with this algorithm.

The computational complexity of this algorithm is: For the forward Fourier transformation,
$\ONotSym(m \cdot n)$ exponentials and $\ONotSym(N \cdot m \cdot n)$ complex multiplications and additions.
For the inverse Fourier transformation, $\ONotSym(N \log(N))$ complex multiplications and additions.

\myheading{Numerical example}
The presented example (Fig. 1) shows the algorithm result in a low $n$ setting ($n=20$),
with observations $X_i$ drawn from a uniform distribution in $[0,19]$. The
average of the sample turned out to be $8.0613 = 40.3066/5$, its minimum and maximum were $0.3872$ and $18.9123$. 
The considered random variable $Z$ was defined as $Z = \sum_{j=1}^5 X^{[j]}$, i.e. $m=5$, 
where the five summands were each deemed independently distributed according to the empirical 
distribution of the sample. Clearly $Z$ takes values in $[1.9360,94.5615]$. Figure 1 
shows the algorithm output when run once with $N=1000$ and once with $N=40$. 
(Note that the true probability density function consists of up to $n^m = 3.2$ million 
singular peaks, or, more rigorously, the distribution function of as many steps.) 

\begin{figure}[htp]
\begin{center}
\subfloat[]{\includegraphics[width=7.2cm\relax]{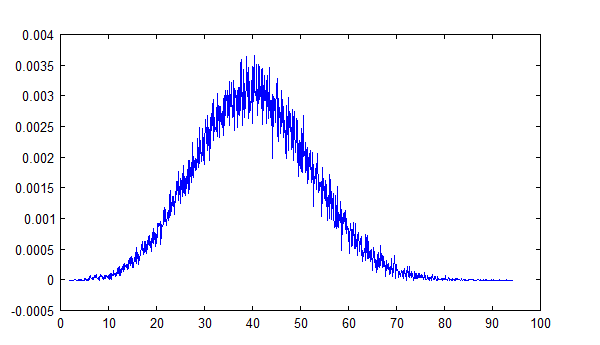}}
\hfill
\subfloat[]{\includegraphics[width=7.2cm\relax]{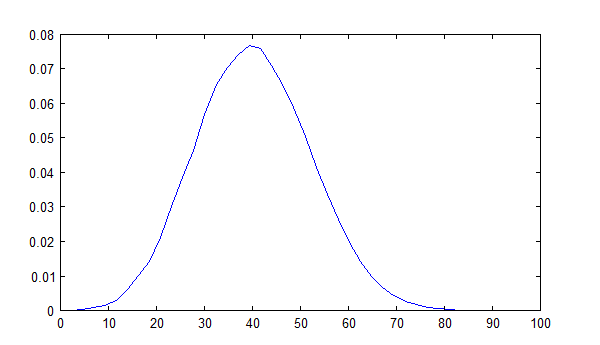}}
\label{fig:test6_2019_03_19b_n20_N1000_m5_unifplusnoise_ones5_T1TZ_truedensity}
\caption{\small{Algorithm output ($f_i$) for an example with sample size $n=20$ 
and mixing width $m=5$ (with $a=\vec 1$), with $N=1000$ (a) and $N=40$ (b).
}}
\end{center}
\end{figure}

\myheading{Conclusion}
Further work should illuminate the convergence characteristics (especially in low $n$/low $m$ settings) in 
dependence on the properties of the input distribution $\hat F_n$.


\appendix
\label{sec:ExamplesOfLinearBootstrapMethods}

\myheading{Appendix: Examples of linear bootstrap methods}\\
\emph{Sample mean in the iid. observations case}\\
The bootstrap estimator $T_n^*$ obtained from centering and norming the 
plugin-estimator of the sample mean, combined with Efron's bootstrap distribution \cite{Efron1979BootstrapMethodsAnotherLook}, 
is a linear boostrap method:
\begin{align} 
    & T_n^* = T(X_1^*, \dots, X_n^*) = \sqrt{n} \cdot ( \tfrac{1}{n} \sum_{i=1}^n X_i^* - (n^{-1} \sum_t X_t) ) \\[-0mm]
    & \text{with } X_i^* \text{ independent and } P^*(X_i^* = X_j) = \tfrac{1}{n}.
\end{align}
\vspace{-2mm}
\mbox{ }\\
\emph{Sample mean for $m$-dependent time series}\\
Let $X_1, \dots, X_n$ be a sample from a strongly stationary $m$-dependent time series. 
(definition: see \cite{Shao1995TheJackknifeAndBootstrap})
Let $n = b \cdot l$, with $l$ the block lengths chosen $> m$, and $b$ denoting the 
number of blocks. The moving block bootstrap estimator for the mean of the time series,
given by 
\begin{align}
    T_n^* & = T(X_1^*, \dots, X_n^*) = \sqrt{b} \cdot (\tfrac{1}{b} \sum_{k=1}^b V_k^* ) \quad \text{ with } \\[-3mm]
    V_k^* & = l^{-1} \cdot \sum_{s=1}^l X_{(k-1)l + s}^* - \bar{\bar X}_n
\end{align}
(where $\bar{\bar X}_n$ is the average of the averages of the $(n-l+1)$ possible consecutive 
blocks in the given time series $X_1, \dots, X_n$; and the $X_i^*$ are drawn according to
the K\"unsch procedure \cite{HansKunsch1989TheJackknifeAndTheBootstrapForGenStatObs}, sec. 2.3) 
essentially is a linear mixture of $b$ independent random variables each taking 
values (with equal probability mass) from the set 
$\{l^{-1} \cdot \sum_{s=1}^l X_{(i-1) + s} - \bar{\bar X}_n, i=1\dots n-l+1\}$.

\vspace{-1mm}
\mbox{ }\\
\emph{Linear regression with iid. noise and fixed design matrix}\\
Details on the linearity property of the residual bootstrap for linear regression estimation
will be elaborated in another text.

\bibliography{dbt_bib}{}
\bibliographystyle{alpha}


\end{document}